\documentclass[11pt]{article}
\pdfoutput=1

\usepackage{euscript}
\usepackage{amssymb}
\usepackage{amsfonts}
\usepackage{amsbsy}
\usepackage{epsfig}
\usepackage{amsthm}
\usepackage{amscd}
\usepackage{amstext}
\usepackage{verbatim}
\usepackage{amsmath}
\usepackage{cancel}
\usepackage{capt-of}
\usepackage{empheq}
\usepackage{subfigure}
\usepackage{xcolor}

\usepackage{cite}
\usepackage{bm}
\usepackage{authblk}
\usepackage[T1]{fontenc}

\usepackage[pdftex,linktoc=all]{hyperref}
\hypersetup{ 
colorlinks=true, 
linkcolor=black, 
citecolor=red, 
}

\def\ben{\begin{equation}}
\def\een{\end{equation}}

\def\be{\begin{equation}}
\def\ee{\end{equation}}
\def\beq{\begin{equation}}
\def\eeq{\end{equation}}
\def\ba{\begin{array}}
\def\ea{\end{array}}

\def\dalemb#1#2{{\vbox{\hrule height .#2pt
       \hbox{\vrule width.#2pt height#1pt \kern#1pt
               \vrule width.#2pt}
       \hrule height.#2pt}}}

\newcommand{\bea}{\begin{eqnarray}}
\newcommand{\eea}{\end{eqnarray}}

\makeatletter
\newcommand*\bigcdot{\mathpalette\bigcdot@{.5}}
\newcommand*\bigcdot@[2]{\mathbin{\vcenter{\hbox{\scalebox{#2}{$\m@th#1\bullet$}}}}}
\makeatother

\renewcommand{\eqref}[1]{(\ref{#1})}

\def\ocal{{\mathcal{O}}}

\title{Phonons, electrons and thermal transport in\\
Planckian high T$_c$ materials}
\author{Connie H. Mousatov and Sean A. Hartnoll
\\ {\it Department of Physics, Stanford University,} \\ {\it Stanford, CA 94305-4060, USA} 
}

\begin{document}
\maketitle

\begin{abstract}

The room temperature thermal diffusivity of high T$_c$ materials is dominated by phonons. This allows the scattering of phonons by electrons to be discerned. We argue that the measured strength of this scattering suggests a converse Planckian scattering of electrons by phonons across the room temperature phase diagram of these materials. Consistent with this conclusion, the temperature derivative of the resistivity of strongly overdoped cuprates is noted to show a kink at a little below 200 K that we argue should be understood as the onset of a high temperature Planckian $T$-linear scattering of electrons by classical phonons. This kink continuously disappears towards optimal doping, even while strong scattering of phonons by electrons remains visible in the thermal diffusivity, sharpening the long-standing puzzle of the lack of a feature in the $T$-linear resistivity at optimal doping associated to onset of phonon scattering.

\end{abstract}

\section{Background}

The thermal conductivity of conventional metals at room temperature obeys the Wiedemann-Franz law \cite{Kumar1993}. This establishes that heat transport is dominated by electrons and that the electronic scattering is elastic. Indeed, at these temperatures electrons are scattered elastically off classical phonons, leading to a $T$-linear resistivity with an underlying scattering rate of $1/\tau \approx k_B T/\hbar$ \cite{P2}. This rate has come to be known as `Planckian' \cite{Zaanen2004}.

The thermal conductivity of unconventional metals such as high $T_c$ cuprates and pnictides at room temperature does not obey the Wiedemann-Franz law \cite{PhysRevB.49.9073,PhysRevB.68.220503,PhysRevB.92.020507, PhysRevB.86.180502,PhysRevB.72.054508}. The Lorenz ratio is larger than the Sommerfeld value, $L > L_0$, often by a factor of 3 or more. Heat transport at room temperature in these materials is dominated by phonons rather than electrons. Nonetheless, as in conventional metals, over important parts of the phase diagram the resistivity is $T$-linear with a Planckian scattering rate \cite{PhysRevB.39.6571, PhysRevB.42.6342,Valla2110, Marel2003,Bruin804, Hartnoll2015}. As we shall see in \S\ref{sec:corr} below, even away from optimal doping the room temperature resistivity has a Planckian $T$-linear component. At these temperatures phonons have every right, as in conventional metals, to be the cause of this Planckian scattering (or, at least, to contribute to it significantly).

There are good reasons to doubt the role of phonon scattering in high T$_c$ electrical transport, articulated forcefully in \cite{10.21468/SciPostPhys.6.5.061}. Firstly, in many high T$_c$ materials Planckian scattering continues to very low temperatures \cite{Legros2019}, where phonons are presumably irrelevant (even allowing for the somewhat small Fermi surfaces of these materials, cf. \cite{PhysRevB.99.085105}). Secondly, $T$-linear resistivity continues to very high temperatures \cite{PhysRevLett.59.1337, PhysRevB.41.846, PhysRevLett.69.2975}, where the short mean free path of the electrons would seem to invalidate a na\"ive scattering picture \cite{PhysRevLett.74.3253}.  Thirdly, the temperature and doping dependence of the resistivity across the phase diagram of high T$_c$ materials --- as well as the behavior of many other observables --- reflects the importance of  electronic correlations \cite{PhysRevLett.93.267001}.
A complete picture must include strong electronic correlations also, although this fact in itself does not negate a potentially important role for phonons.

In this paper we make two observations concerning the role of electron-phonon scattering in these materials. We will be interested in intermediate temperatures, where (some) phonons are classical but where electronic mean free paths are not yet extremely small.
Firstly, in \S\ref{sec:react}, we argue that at these temperatures the scattering of phonons by electrons is visible in the large phonon contribution to heat transport in high T$_c$ materials. This gives direct evidence for the occurrence of strong electron-phonon interactions. These same interactions are then argued to lead, conversely, to a Planckian lifetime for electrons due to scattering by phonons. Secondly, in \S\ref{sec:corr}, we note that the temperature derivative of the resistivity of heavily overdoped cuprates shows a kink at the temperature scale where scattering of electrons by classical phonons is expected to onset. We argue that this, again, demonstrates the occurrence of strong electron-phonon interactions. The kink is seen to disappear continuously towards optimal doping, offering a new angle of approach to the long-standing puzzle of the lack of features in the optimally doped resistivity due to phonon scattering.

Our emphasis on the importance of electron-phonon scattering processes for transport in high-T$_c$ materials at temperatures above around 200 K builds on the measurements and interpretation of thermal diffusivity in \cite{Zhang5378, PhysRevB.100.241114, Zhang19869}. Here we show how the experimental results can be understood within a quasiparticle picture of electron-phonon interactions. Furthermore, in \S\ref{sec:highTc} we extend the analysis of \cite{PhysRevB.100.241114} to different high T$_c$ materials and dopings, using existing data for the thermal conductivity, specific heat and electrical resistivity. And in \S\ref{sec:disorder} we establish a parallel between high temperature thermal and electric transport in high T$_c$ materials and in heavily doped semiconductors. 
The role of electron-phonon scattering has been well-characterized in the semiconductors, supporting our general discussion.

\section{Action and reaction of electrons and phonons}
\label{sec:react}

Our starting point is recent measurements of the thermal diffusivity $D_\text{th}$ in several cuprates near optimal doping at temperatures $T$ up to 600 K \cite{Zhang5378, PhysRevB.100.241114}. Above around 200 K the inverse thermal diffusivity, a measure of thermal resistivity, is found to behave as
\begin{equation}\label{eq:Dth}
D_\text{th}^{-1} = \frac{\lambda \, T}{v_s^2} + D_0^{-1} \,, 
\end{equation}
with the temperature-independent coefficients
\begin{equation}\label{eq:c1c2}
\lambda \sim \lambda_o \frac{k_B}{\hbar} \,, \qquad D_0^{-1} \sim \frac{m_\star}{\hbar} \,. 
\end{equation}
Here $v_s$ is the sound speed and $m_\star$ the effective electron mass. The dimensionless prefactor $\lambda_o \sim 1/5$ in these materials \cite{PhysRevB.100.241114}.
The leading high temperature behavior $D_\text{th} \sim v_s^2 \times \hbar/(k_B T)$ in (\ref{eq:Dth}) is characteristic of crystalline insulators with poor thermal conduction \cite{Zhang19869, Behnia_2019, Mousatov2020}. This term is due to the degradation of the phonon heat current due to phonon anharmonicity and is entirely analogous to the $T$-linear scattering of electrons by classical phonons. The fact that the high temperature heat transport of unconventional metals mirrors that of insulators is consistent with the dominant heat transport by phonons seen in the large Lorenz ratio, mentioned above.

The constant offset term in (\ref{eq:Dth}), however, is not present in crystalline insulators \cite{Zhang19869}. We have already seen that the heat is carried by phonons. The inverse diffusivity is thus proportional to the phonon scattering rate.
The additive inverse diffusivity in (\ref{eq:Dth}) therefore immediately suggests a Matthiessen rule in which a single degree of freedom undergoes two distinct scattering mechanisms. Given that the constant term is absent in insulators, the most natural additional scattering process is that of phonons by charge carriers (more precisely, particle-hole pairs). That is, it is natural to read (\ref{eq:Dth}) as
\begin{equation}\label{eq:diff}
D_\text{th}^{-1} \sim D_\text{th,ph}^{-1} \sim \frac{d}{v_s^2} \frac{1}{\tau_{\text{ph}}} = \frac{d}{v_s^2} \left( \frac{1}{\tau_{\text{ph} \to \text{ph}}} + \frac{1}{\tau_{\text{ph} \to \text{el}}} \right) \,,
\end{equation}
where scattering rates have been added and $d$ is the number of spatial dimensions. From the comments below (\ref{eq:c1c2}) and with $d=3$ one has $\tau^{-1}_{\text{ph} \to \text{ph}} \sim \frac{1}{15} k_B T/\hbar$. The fact that the phonons have longer lifetimes than the Planckian electrons is important for the phonons to be able to dominate thermal transport, as is the fact that the Fermi velocity is relatively small in these systems ($v_F/v_s \approx 35$ \cite{PhysRevB.100.241114}). Let us now discuss $\tau_{\text{ph} \to \text{el}}^{-1}$.

It is a fundamental physical principle that to every action there is an equal and opposite reaction. This logic will connect the timescale $\tau_{\text{el} \to \text{ph}}$ for an electron to emit and re-absorb a phonon with the timescale $\tau_{\text{ph} \to \text{el}}$ for a phonon to decay into a particle-hole pair. These two timescales need not themselves be equal but they are both controlled by the strength of the coupling between electrons and phonons. When 
a dimensionless measure of the electron-phonon coupling (the deformation potential relative to the characteristic energy scales of the electrons and phonons) is order one in magnitude, so that the electronic lifetime $\tau_{\text{el} \to \text{ph}}$ is Planckian, the phonon lifetime $\tau_{\text{ph} \to \text{el}}$ takes a characteristic value. We show in the Supplementary Material that, in agreement with well-established results for electron-phonon scattering \cite{zim1},
\begin{align}\label{eq:planck}
\frac{1}{\tau_{\text{el}\rightarrow \text{ph}}} \sim \frac{k_{B}T}{\hbar} \qquad \Leftrightarrow \qquad \frac{1}{\tau_{\text{ph}\rightarrow \text{el}}} \sim \frac{m_\star v_s^2}{\hbar} \,.
\end{align}
We have assumed that the electrons are degenerate with a large Fermi surface so that the Fermi wavelength is comparable to the lattice spacing, $k_F a \sim 1$, while the phonons are classical. We have neglected all numerical factors, that depend on microscopic considerations. The energy scale $m_\star v_s^2$ in (\ref{eq:planck}) arises as the Debye energy squared over the Fermi energy: $\tau^{-1}_{\text{ph} \to \text{el}} \sim \hbar \omega_D^2/E_F \sim m_\star v_s^2/\hbar$, where $E_F \sim \hbar^2/(m_\star a^2)$ and  $\omega_D \sim v_s/a$. The result (\ref{eq:planck}) also assumes that the phonon system can self-equilibrate on faster timescales than it returns momentum and energy to the charge carriers (Bloch's {\it Annahme}). At high temperatures this is reasonable due to the strong effects of lattice anharmonicity ($\tau^{-1}_{\text{ph} \to \text{ph}} \gtrsim \tau^{-1}_{\text{ph} \to \text{el}}$).

Using the second relation in (\ref{eq:planck}) in the diffusivity (\ref{eq:diff}) recovers precisely the observed offset in (\ref{eq:Dth}) and (\ref{eq:c1c2}).
Therefore the single assumption of electron-phonon scattering with an order one dimensionless coupling constant simultaneously leads to --- at these high temperatures ---  both the observed Planckian charge transport and the offset to the observed Planckian heat transport. This is not a strong assumption, and holds in conventional metals.

\section{Thermal diffusivity in high T$_c$ materials}
\label{sec:highTc}

Here we extend the analysis of thermal diffusivity in \cite{PhysRevB.100.241114} to different families of high T$_c$ materials and different dopings. We use existing thermal conductivity $\kappa$ and specific heat $c$ data for those materials. The thermal diffusivity is then given by $D_\text{th} = \kappa/c$. The direct thermal diffusivity measurements in \cite{Zhang5378, PhysRevB.100.241114} extended in several cases to 600 K. This allowed the regime of behavior (\ref{eq:Dth}) to be clearly visible. We are not aware of any other published measurements of both $\kappa$ and $c$ in metallic high T$_c$ compounds up to these temperatures. Our analysis is therefore limited to temperatures up to room temperature. The onset of the behavior (\ref{eq:Dth}) is visible in the temperature range of 200 to 300 K, but the linearity of $D_\text{th}^{-1}$ in $T$ is typically not fully developed. The diffusivity below 200 K remains dominated by phonons, but the more complicated temperature dependence means that scattering by charge carriers and by other (no longer classical) phonons are less easily separated out. In our opinion, further measurements to higher temperatures are highly desirable.

\begin{figure}[ht!]
    \centering
    \includegraphics[width=\textwidth]{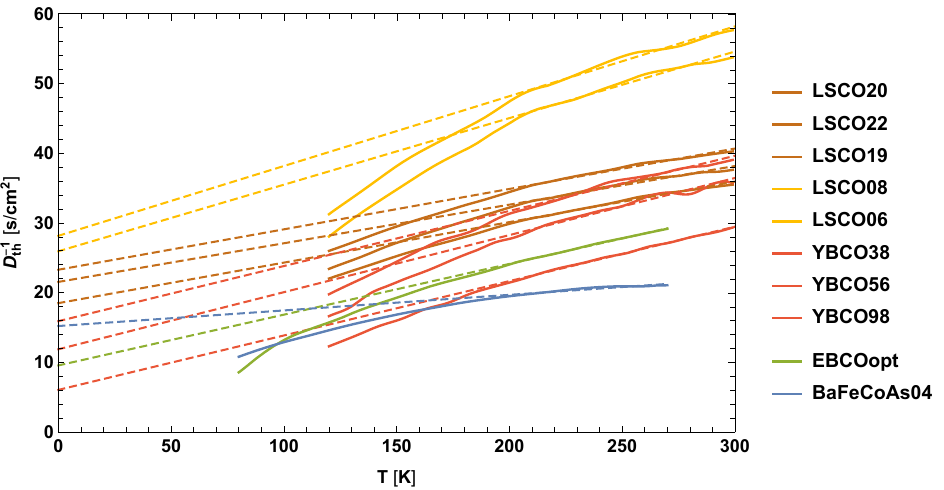}
    \caption{Inverse thermal diffusivity for several high T$_c$ compounds. Solids lines are data and the dashed lines show the intercept of the high temperature linear in $T$ regime. Within each group (overdoped and underdoped LSCO \cite{Yan2004,LSCOa}, $a$-axis YBCO \cite{PhysRevB.68.220503, LORAM1990243}, EBCO \cite{ PhysRevB.72.054508, BEDNARZ1990385} and Ba-122 \cite{PhysRevB.92.020507,PhysRevB.79.054525,PhysRevB.88.144502}) the curves are in the order shown in the legend.}
    \label{fig:highTc}
\end{figure}

Fig.~\ref{fig:highTc} shows $D_\text{th}^{-1}$ for several materials including underdoped LSCO and YBCO, overdoped LSCO and close-to-optimally doped EBCO and YBCO. We have restricted attention to compounds where electrical resistivity measurements in the same (or very similar) crystals over the same temperature range show that the Lorenz ratio is large (see Supplementary Material). This fact suggests that phonons dominate the heat current.\footnote{We will not attempt to further isolate the phonon heat current by subtracting out the electronic contribution using the Wiedemann-Franz law. Such subtractions are imprecise, even in conventional systems, and furthermore the extracted values of the intercept $D_0^{-1}$ are not found to change significantly.} Most of the variation seen in $D_\text{th}^{-1} = c/\kappa$ comes
from the specific heat, while the thermal conductivity is almost constant. This highlights the fact that $D_\text{th} \sim v_s^2 \tau_\text{ph}$ is more physically transparent than $\kappa$ here: the temperature variation of $D_\text{th}^{-1}$ directly relates to the phonon scattering rate, showing the onset of scattering by classical phonons above 200K.

Fig.~\ref{fig:highTc} also shows the extrapolation of the high temperature linear in $T$ regime to obtain the intercept $D_0^{-1}$. Due to the limited temperature range of data available the linear regime is not fully developed. In the data of \cite{PhysRevB.100.241114} this regime is seen to onset at around 200 K. We have therefore made the linear fits in the temperature range of 200 to 300 K. The intercepts are seen to lie in the range of around 10 to 30 s/cm$^2$, consistent with the values obtained from higher temperature diffusivity measurements in \cite{PhysRevB.100.241114, Zhang19869}.

We now argue that the intercepts in Fig.~\ref{fig:highTc} indicate a scattering rate $\tau^{-1}_{\text{ph}\to\text{el}} = \ocal(1) \times m_\star v_s^2/\hbar$. This is in agreement with the estimates for other materials in \cite{PhysRevB.100.241114} and compatible with Planckian charge carriers according to the `action-reaction' relation (\ref{eq:planck}).
The starting point is the fundamental quantity $m_e/\hbar \approx 0.86$ s/cm$^2$. The larger 
intercepts in Fig.~\ref{fig:highTc} are due to two additional factors. Firstly, the factor of $d \sim 3$ in the diffusivity (\ref{eq:diff}); heat diffusion is three dimensional but anisotropic. Secondly, the enhancement of the effective mass $m_\star$. In YBCO, quantum oscillations and optical measurements suggest $m_\star \approx 2 - 3 m_e$ \cite{PhysRevB.72.060511, Doiron-Leyraud2015} while in underdoped LSCO the effective mass is larger \cite{PhysRevB.72.060511}, and, from specific heat measurements, significantly larger on the overdoped side: $m_\star \approx 5 - 10 m_e$ \cite{Legros2019}. These mass renormalizations lead to comparable values of $D_0^{-1}\times \hbar/m_\star$ for the different compounds in Fig.~\ref{fig:highTc}.

Over the temperature range where the thermal diffusivity is fit, the resistivity is metallic but is only precisely $T$-linear for the optimally doped samples \cite{PhysRevLett.93.267001}.
Over this range the product of the Fermi momentum and electronic mean free path as extracted from the resistivity, $k_F \ell = h/e^2 \times d_c/\rho$ with $d_c$ the spacing between conducting planes, is greater than $2 \pi$ for the overdoped samples but slightly less than $2 \pi$ for the underdoped samples. A simple quasiparticle picture of charge transport may only be marginally valid in these latter cases.

The curve for the close-to-optimally electron-doped pnictide Ba-122 is noticeably flatter than the others. This high temperature behavior is suggestive of a saturated `glassy' phonon mean free path $\ell_\text{ph}$ rather than the $\ell_\text{ph} \sim 1/T$ characteristic of crystalline phonons, as has been reported previously in insulating BSYCO \cite{PhysRevB.49.9073}. Saturation may be due to strong disorder \cite{PhysRevB.46.6131} and/or strong lattice anharmonicity \cite{PhysRevB.82.224305}.

\section{Disorder and comparison to heavily doped semiconductors}
\label{sec:disorder}

The main confounding factor for our proposed interpretation of the thermal diffusivity data in terms of phonon scattering by charge carriers is the role of disorder. As we explain below, scattering of phonons by disorder also contributes to the offset $D_0^{-1}$. 
Doping introduces both disorder and charge carrier scattering simultaneously, so a more refined analysis is needed to tease these effects apart. Here we argue that the offset due to disorder scattering does not overwhelm that due to scattering by charge carriers.

The scattering of phonons by both disorder and charge carriers has been extensively studied in the thermal transport of heavily doped semiconductors. Heavily doped semiconductors have similarities with both high T$_c$ compounds and conventional materials: high temperature thermal transport is dominated by phonons and the inverse diffusivity obeys (\ref{eq:Dth}), with an offset comparable to that in high T$_c$ materials, while electrical resistivity is $T$-linear and Planckian, due to scattering of degenerate electrons by phonons. In the Supplementary Material we review experimental results in heavily doped Ge-Si alloys \cite{dis, ger, PhysRev.136.A1149} as well as heavily doped single silicon crystals \cite{doi:10.1002/aelm.201600171}.

The essential point emerging from the understanding of heavily doped semiconductors is that while mass disorder effectively scatters high frequency phonons, it does not scatter low frequency phonons as efficiently because the scattering needs to induce vibrations in the impurity atoms \cite{Klemens_1955}. The surviving long wavelength phonons that do transport heat are especially sensitive to scattering by charge carriers, because they are able to excite particle-hole pairs on the Fermi surface. This allows phonon-electron scattering to have a significant effect on phonon heat transport \cite{PhysRev.136.A1149, PhysRevLett.114.115901, doi:10.1063/1.5108836}.

The frequency dependence of scattering is captured by the widely employed Callaway model for thermal transport by phonons \cite{PhysRev.113.1046,PhysRev.120.1149}, extended to include phonon-electron scattering \cite{PhysRev.136.A1149}. This model has previously been used to estimate the role of disorder and carrier scattering in the Ge-Si alloys \cite{PhysRev.136.A1149} and doped silicon \cite{doi:10.1002/aelm.201600171} samples that we consider in the Supplementary Material. At high temperatures the model gives the (three dimensional) thermal diffusivity as
\be\label{eq:call}
D_\text{th,ph} =  \frac{v_s^2}{3}\int_0^{1} \hat \omega^2 \tau_\text{ph}(\hat \omega) d\hat \omega = \frac{v_s^2}{3} \int_0^{1} \frac{\hat \omega^2 d\hat \omega}{\tau^{-1}_\text{ph,dis} \, \hat \omega^4 + \tau^{-1}_{\text{ph} \to \text{el}} \, \hat \omega + \tau^{-1}_{\text{ph} \to \text{ph}} \, \hat \omega^2} \,.
\ee
Here $\hat \omega = \omega/\omega_D$, with $\omega_D$ the Debye scale, and the denominator in the final expression contains three terms corresponding to scattering from disorder, charge carriers and other phonons. The factor of $\hat \omega^2$ in the numerator comes from the phase space for acoustic phonons. 
The inverse timescales that appear in the denominator are frequency-independent. The quantities $\tau^{-1}_{\text{ph} \to \text{el}}$ and $\tau^{-1}_{\text{ph} \to \text{ph}}$ have been given previously in the text. The timescale for scattering of phonons by disordered impurities is
\be\label{eq:taudis}
\frac{1}{\tau_{\text{ph,dis}}} \sim x \, \omega_D \,.
\ee
Here $x$ is the density of impurities together with other factors proportional to the change of the local atomic vibrational energy due to the substitution of the atom (e.g. \cite{Klemens_1955,PhysRev.136.A1149}).

In the asymptotic high temperature limit phonon scattering dominates so that $\tau^{-1}_{\text{ph} \to \text{ph}} \gg \tau^{-1}_\text{ph,dis}, \tau^{-1}_{\text{ph} \to \text{el}}$. In this limit the integral in (\ref{eq:call}) can be performed explicitly to yield the inverse diffusivity
\be\label{eq:withdisorder}
D_\text{th,ph}^{-1} = \frac{3}{v_s^2} \left(\frac{1}{\tau_{\text{ph} \to \text{ph}}} + \frac{1}{12} \frac{1}{\tau_\text{ph,dis}} + \frac{1}{\tau_{\text{ph} \to \text{el}}} \log \frac{\tau_{\text{ph} \to \text{el}}}{\tau_{\text{ph} \to \text{ph}}} + \ocal(\tau_{\text{ph} \to \text{ph}}) \right) \,.
\ee
This expression amounts to a refined version of (\ref{eq:diff}). Relative to an estimate that neglects frequency dependence, scattering from carriers is logarithmically enhanced while disorder scattering is suppressed by a factor of $1/12$. This suppression is due to the fact that low frequency phonons are not efficiently scattered by atomic impurities. If follows from (\ref{eq:withdisorder}) that the relative contribution of disorder and carrier scattering to the high temperature offset $D_0^{-1}$ is determined, for sufficiently large Fermi surfaces, by
\be\label{eq:comp}
\frac{x}{12} \omega_D \qquad vs. \qquad \frac{\hbar \omega_D}{E_F} \omega_D \,.
\ee
Here we neglected the logarithmic enhancement and recalled that 
the estimate (\ref{eq:planck}) came from $\tau^{-1}_{\text{ph} \to \text{el}} \sim \hbar \omega_D^2/E_F \sim m_\star v_s^2/\hbar$. In the regime where phonons are strictly classical and charge carriers degenerate, $\hbar \omega_D \ll k_B T \ll E_F$, the carrier scattering contribution in (\ref{eq:comp}) is small because $\hbar \omega_D/E_F \ll 1$. However, in the heavily doped semiconductors the ratio $\hbar \omega_D/E_F$ is not extremely small, larger than 1/5, while $x$ is order one for the alloys and at least an order of magnitude smaller for the doped silicon samples. The contribution due to scattering by electrons is not overwhelmed. Indeed, the importance of carrier scattering in these compounds is corroborated by more thorough first principles simulation \cite{PhysRevLett.114.115901, PhysRevLett.106.045901, doi:10.1063/1.5108836}.

While the phonon band structure is more complicated in cuprates (c.f the discussion of complex oxides in \cite{Zhang19869, Mousatov2020}), the comparison (\ref{eq:comp}) gives a rough estimate of the relative importance of disorder and carrier scattering. The disorder caused by doping in cuprates is in between that of the silicon samples and the Ge-Si alloys while $\hbar \omega_D/E_F$ is only somewhat smaller in cuprates. Thus (\ref{eq:comp}) supports our claim that phonon-carrier scattering makes a significant contribution to the offset of the high temperature inverse diffusivity in cuprates.

\section{Phonons and overdoped cuprates}
\label{sec:corr}

We have argued that the offset of the high temperature inverse thermal diffusivity in Fig.~\ref{fig:highTc} (and in \cite{PhysRevB.100.241114}) is, via the relation (\ref{eq:planck}), suggestive of Planckian scattering of electrons by phonons across the phase diagram above about 200 K. To see this scattering directly in electronic transport, we look to strongly overdoped samples. These are believed to have more conventional dynamics and, in particular, different scattering mechanisms can be expected to be additive.
In this section we see that the resistivity of sufficiently overdoped samples indeed shows the onset of an additive Planckian $T$-linear term at these temperatures.

\begin{figure}[ht!]
    \centering
    \includegraphics[width=0.49\textwidth]{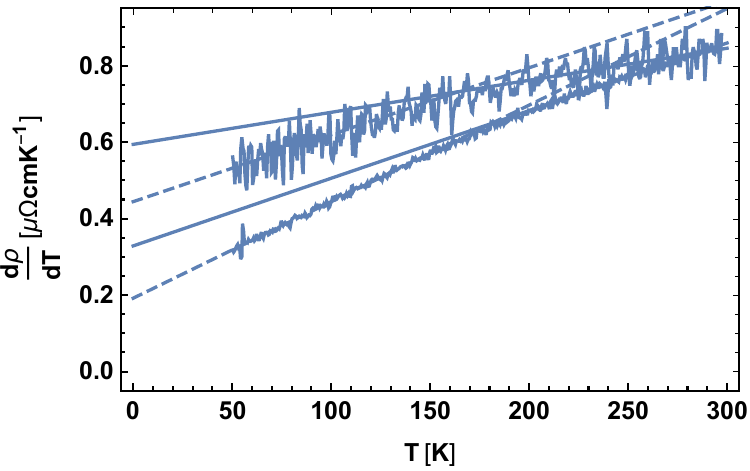}\includegraphics[width=0.49\textwidth]{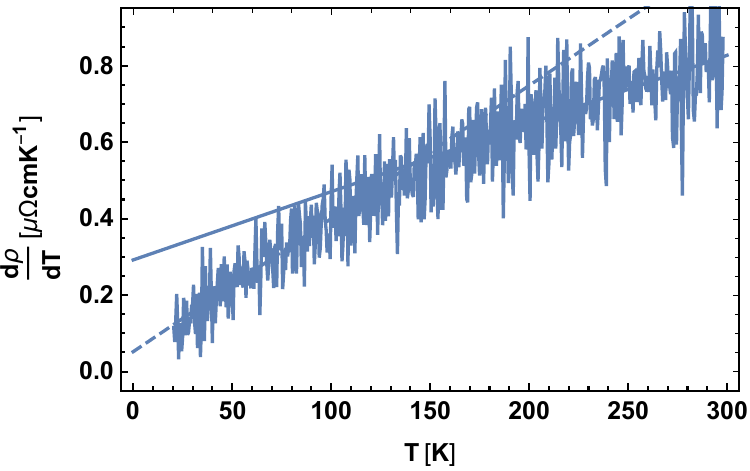}
    \caption{Temperature derivative of the resistivity for overdoped LSCO. Left plots have $p=0.26$ (top) and $p=0.29$ (bottom). Right plot has $p=0.33$. Data from \cite{dich}.
    Dashed lines are fits to $a_o + 2 b_o T$ at low temperatures (as in \cite{dich}), while solid lines are fits to $a + 2 b T$ in the temperature range of 200 to 300 K.}
    \label{fig:over}
\end{figure}

Fig.~\ref{fig:over} shows that the resistivity of heavily overdoped LSCO (doping $p=0.26, 0.29$ and $0.33$) at around room temperature is of the form $\rho = \rho_0 + a T + b T^2$. A kink in the temperature derivative of the resistivity is clearly discerned at a little below 200 K. The kink implies that the room temperature regime is not the same as the low temperature `anomalously critical' resistivity, as analyzed in e.g. \cite{PhysRevB.53.5848, Cooper603}. The persistence of a $T^2$ term indicates that this is also not the purely $T$-linear `incoherent' regime of e.g. \cite{dich}. The room temperature $T+T^2$ behavior extends to very high temperatures at these dopings \cite{PhysRevLett.69.2975}. Given that 200 K is (from e.g. our Fig.~\ref{fig:highTc} and also plots of $d\rho/dT$ from Bloch-Gruneisen theory \cite{Pickett1991}) the scale at which $T$-linear phonon scattering onsets in cuprates, it is natural to understand the room temperature 
regime as phonon $T$-linear plus electronic $T^2$ scattering. The increase in the offset in Fig.~\ref{fig:over} above the kink indicates an additional $T$-linear scattering mechanism at high temperatures (classical phonons, we are claiming). 
The simultaneous decrease in the slope indicates a reduction in the $T^2$ scattering.
This latter scattering is presumed to be electronic and the reduction may be due to the expected decrease of the electronic density of states upon crossing the phonon energy scale.

\begin{figure}[ht!]
    \centering
    \includegraphics[width=0.75\textwidth]{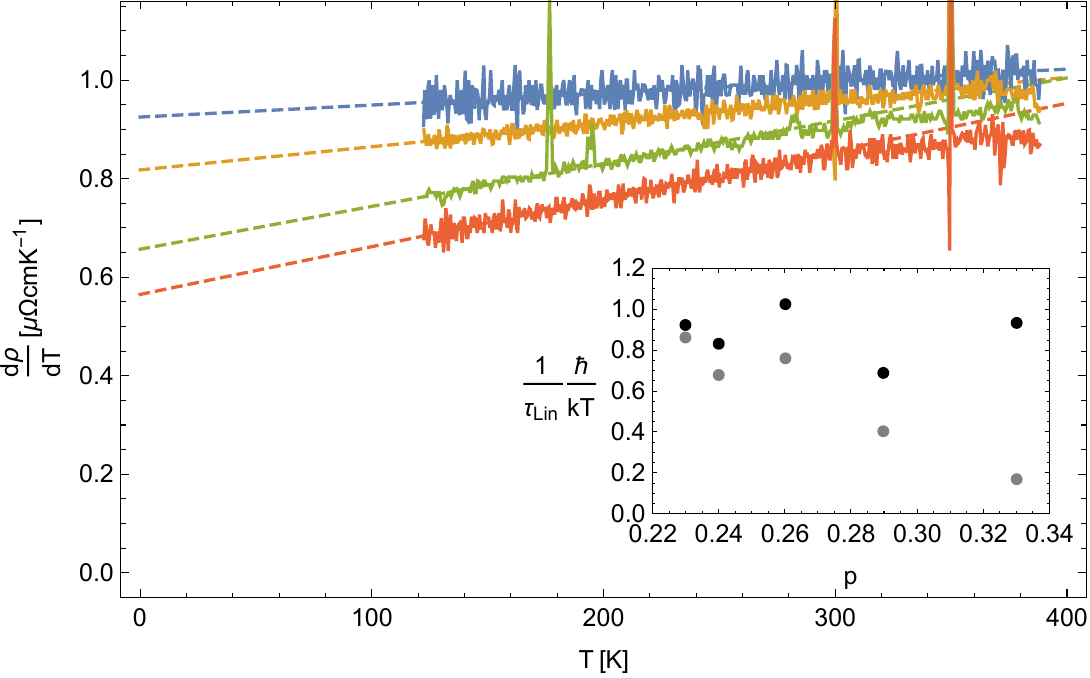}
    \caption{Temperature derivative of the resistivity for mildly overdoped LSCO. From top to bottom $p=0.19, 0.20, 0.21, 0.22$.
    Data from \cite{PhysRevLett.93.267001}. Dashed lines are fits to $a + 2 b T$. The inset shows the merging of the higher temperature (upper, black dots) and lower temperature (lower, gray dots) $T$-linear scattering rates as doping is reduced. Data for the inset from \cite{dich}, with the timescale expressed in Planckian units following \cite{Legros2019}.
    }
    \label{fig:lessover}
\end{figure}

Fig.~\ref{fig:over} shows two distinct regimes of $\rho \sim T + T^2$ behavior. In the low temperature regime the coefficient of the $T$-linear term in the resistivity (hence, the constant offset of the derivative) decreases with overdoping and becomes small at around $p = 0.33$, as discussed in \cite{Cooper603} and shown in the inset of Fig.~\ref{fig:lessover}. In the higher temperature regime (fit between 200 and 300 K) the $T$-linear term is instead roughly constant and saturates away from zero at large doping. This is also shown in the inset of Fig.~\ref{fig:lessover}. Using the estimates in \cite{Legros2019} for the effective mass and density of carriers at these dopings, the scattering timescale associated to the $T$-linear term in the high temperature regime is found to be Planckian, with $\tau^{-1} \approx \{1, 0.7, 0.9\} \times k_B T/\hbar$ for these highest three dopings, respectively. These rates are  consistent with the expected Planckian scattering of electrons by phonons.

The samples in Fig.~\ref{fig:over} are significantly more overdoped than those whose thermal transport was considered in Fig.~\ref{fig:highTc}. These more overdoped samples will have a larger electronic contribution to thermal transport and the analysis we performed above may not be directly applicable. The temperature derivative of the resistivity of less overdoped samples is shown in Fig.~\ref{fig:lessover}, with $p=0.19, 0.20, 0.21, 0.22$.
Remarkably, at these lower dopings there is no kink, and the low temperature $\rho \sim T + T^2$ behavior continues up to around 300K or higher. The inset of the figure shows how the kink in the $T$-linear scattering rate --- which remains fixed at a little below 200K --- disappears as doping is reduced. At the highest temperatures in Fig.~\ref{fig:lessover} there is a flattening of the derivative associated with proximity to optimal doping \cite{PhysRevLett.69.2975, dich}. This is distinct from the kink at lower temperature that we are discussing.

We have focussed here on overdoped LSCO, which has been the most studied in previous work. However, the same phenomenology --- a feature in $d\rho/dT$ at a little below 200K that disappears at small overdoping --- is also seen in overdoped Bi2201 \cite{putzke2020reduced}.

A long standing puzzle of optimally doped cuprates is why the $T$-linear resistivity fails to show a feature associated to the onset of scattering from classical phonons above some temperature. In Figs. \ref{fig:highTc} and \ref{fig:over} we have seen two instances in which such a feature is, in fact, visible. Fig.~\ref{fig:lessover} furthermore shows how one of the features disappears in mildly overdoped samples, but before optimal doping is reached. This may be a fruitful angle to attack this puzzle. In particular, the presence of an additive $T^2$ term throughout the disappearance may shore up our confidence in a quasiparticle description.\footnote{In contrast, perfectly $T$-linear resistivity at the lowest temperatures, seemingly without a $T^2$ term, has been reported close to a critical doping in e.g. LSCO \cite{Cooper603}, Nd-LSCO \cite{Daou2009}, PCCO \cite{PhysRevLett.81.4720} and LCCO \cite{Jin2011}.}  A possible explanation for the absence of a feature in the resistivity close to optimal doping is a Planckian bound on scattering that prevents scattering channels from adding beyond a certain point \cite{Bruin804, Hartnoll2015}. If a quasiparticle description indeed holds in these mildly overdoped materials at around 200 K, one can hope to understand in detail how non-phonon scattering below 200 K manages to continuously morph into phonon scattering above 200K.

\section{Discussion}

We have argued that recent high temperature (above 200 K) measurements of thermal transport in high T$_c$ materials reveal phonon scattering by charge carriers. As part of this argument we have pointed out that the high T$_c$ thermal diffusivity data is very similar to that in heavily doped semiconductors, where the scattering of phonons by carriers is known to be important. Interpreted in this way, the data determines the strength of electron-phonon interactions in these systems. We have shown that the interaction strength following from the data implies that the converse scattering of charge carriers by phonons leads to a Planckian lifetime for the charge carriers. This in itself is more of a consistency check than a surprise; a Planckian lifetime due to scattering by phonons at these temperatures is ubiquitous in conventional metals (and, we have noted, in heavily doped semiconductors).

However, in high T$_c$ materials electronic correlations are also important at the same high temperature scales. It has been known for some time that charge carriers close to optimal doping have Planckian lifetimes both above and below the temperature scale at which scattering from classical phonons can occur. The slope of the resistivity does not change across that scale.
From this fact one might be tempted to draw the conclusion that electron-phonon scattering has somehow been `de-activated' by, or subsumed into, stronger electronic dynamics. Much theoretical modelling of this regime neglects phonons entirely, as do ultracold atomic simulations of the Hubbard model. From this perspective, scattering at room temperature is expected to be pure electronic close to the critical doping. On the contrary, our interpretation of the thermal diffusivity data, refining the arguments in \cite{Zhang5378, PhysRevB.100.241114, Zhang19869}, suggests that strong electron-phonon scattering does occur and can be seen in transport. This conclusion is consistent with the observation of strong electron-phonon interactions in angle-resolved photoemission at energy scales that are compatible with the transport phenomena we have discussed, e.g. \cite{Zhou2003, He62}. It may be interesting to study this tension within a model that self-consistently incorporates electron-phonon scattering and the onset of electronic quantum criticality.

\section*{Acknowledgements}
We are grateful to J.R. Cooper, N. Hussey and S. Komiya for sending us raw data. We have benefited from discussions with Kamran Behnia, Steve Kivelson, Andy Mackenzie, Akshat Pandey and Veronika Sunko. S.~A.~H. is supported by a Simons Investigator award. C.~H.~M.~ is supported by an NSF graduate fellowship. S.~A.~H. acknowledges the hospitality of the Max Planck Institute CPfS.

\bibliographystyle{ourbst}
\bibliography{refs}

\newpage

\appendix

\section*{Supplementary Material}

\section{Electron and phonon lifetimes}
\label{sec:ratio}

Here we sketch the computation of the electron lifetime due to emission and absorption of a phonon and the phonon lifetime due to interaction with an electronic particle-hole pair. This is textbook material and in particular the phonon lifetime was obtained in \cite{zim1}. We describe the additional assumptions leading to the relationship (\ref{eq:planck}) between the electron and phonon scattering rates.

Our computation is concerned with regimes where it is possible to add scattering rates. This means that we can treat the electron-phonon scattering by starting with non-interacting quasiparticles. The non-interacting electrons and phonons have dispersion $\epsilon_k$ and $\omega_k$, respectively. Their spectral weight is therefore given by
\begin{align}
A(\omega,k) = \pi \delta(\omega - \epsilon_k) \,, \qquad
D(\omega,k) = \pi \left[\delta(\omega - \omega_k)+ \delta(\omega + \omega_k) \right] \,.
\end{align}
Interactions between electrons and phonons then lead to the following lifetimes
\begin{align}
\Sigma_\text{el}''(\omega,p) & = \lambda^2 \int \frac{d^dk}{(2 \pi)^d} \frac{d\Omega}{\pi} \frac{f(\omega-\Omega) b(\Omega)}{f(\omega)} A(\omega-\Omega,p-k) D(\Omega,k)  \,, \label{eq:p1} \\
\Sigma_\text{ph}''(\omega,p) & = \lambda^2 \int \frac{d^dk}{(2 \pi)^d} \frac{d\Omega}{\pi} \frac{f(\omega-\Omega) f(\Omega)}{b(\omega)} A(\omega-\Omega,p-k) A(-\Omega,-k)  \,. \label{eq:p2}
\end{align}
Here $\lambda$ is the electron-phonon coupling, $b(\omega) = 1/(e^{\hbar \omega/k_B T} - 1)$ and $f(\omega) = 1/(e^{\hbar \omega/k_B T} + 1)$.
We work with $d$ spatial dimensions.
We have written these expressions such that the lifetimes are manifestly determined by the available phase space for scattering. At the high temperatures that we will be considering we do not need to keep track of the momentum dependence of the coupling $\lambda$ (which is certainly important at low temperatures).

We can estimate (\ref{eq:p1}) and (\ref{eq:p2}) in the temperature range $\hbar \omega_D \ll k_B T \ll E_F$. Here $\hbar \omega_D$ is the Debye energy, and $E_F$ the Fermi energy. The electrons are therefore degenerate while the phonons are classical. While this separation of scales does not always hold in the experimental regimes, it allows the dependence of the lifetimes on electronic and lattice parameters to be estimated in a transparent way. In particular, the phonon dispersion is bounded by the Debye scale $\omega_k \leq \omega_D$. Therefore at these temperatures $\hbar \omega \sim k_B T \gg \hbar \Omega$ in the integral (\ref{eq:p1}) for the electron self energy, so that
\be
\frac{1}{\tau_{\text{el}\to \text{ph}}} = \Sigma_\text{el}''(\omega,p) = \pi \lambda^2 k_B T \int \frac{d^dk}{(2 \pi)^d} \frac{\delta (\omega - \epsilon_{p-k} \pm \omega_k)}{\hbar \omega_k} \sim \frac{\lambda^2 k_F^{d-1}}{v_F \omega_D} \frac{k_B T}{\hbar} \,. \label{eq:L1}
\ee
In the last step we have estimated $\omega_k \sim \omega_D$. The hierarchy of scales $\hbar \omega_D \ll k_B T \ll E_F$ means that the $k$ integral only gets contributions from close to the Fermi surface. The $k_F^{d-1}$ comes from integrating over the Fermi surface, while the $1/v_F$ is from the perpendicular direction (for which $d k_\perp \sim d\epsilon/v_F$). 

In the phonon self energy (\ref{eq:p2}) we have $\hbar \omega \ll k_B T$, while $\hbar \Omega \sim k_B T$. Therefore
\be
\frac{1}{\tau_{\text{ph}\to \text{el}}} = \Sigma_\text{ph}''(\omega,p) = \frac{\pi \lambda^2 \hbar \omega}{k_B T} \int \frac{d^dk}{(2 \pi)^d} f(\epsilon_{-k}) f(- \epsilon_{-k}) \delta(\omega + \epsilon_{-k} - \epsilon_{p-k}) \sim \frac{\lambda^2 \omega_D k_F^{d-2}}{v_F^2} \,. \label{eq:L2}
\ee
Here the factors of the Fermi-Dirac distribution restrict the energy $\hbar |\epsilon_{-k}| \lesssim k_B T$. This leads to a factor of order $k_B T/(\hbar v_F)$ from the $k$ integral perpendicular to the Fermi surface. The delta function requires both $-k$ and $-k+p$ to be close to the Fermi surface (again because the electronic energy scales are greater than the temperature or lattice energy scales). This leads to one of the remaining integrals being of order $1/v_F$. Finally, we estimated $\omega \sim \omega_D$. The constraint that $-k$ and $-k+p$ both be close to the Fermi surface requires that the phonon wavevector $|p| < k_F$. We have assumed this also. A more complete expression, showing how the scattering turns off for small Fermi surfaces, is given in \cite{zim1}.

The coupling $\lambda$ is related to the deformation potential $D$ by $\lambda^2 a^2 \sim D^2/(\rho_M \hbar \omega_D)$. Here $\rho_M = M/a^d$ is the atomic mass density. More relevant for our purposes is the ratio of lifetimes following from (\ref{eq:L1}) and (\ref{eq:L2})
\be\label{eq:ratio}
\frac{\tau_{\text{ph}\to \text{el}}}{\tau_{\text{el}\to \text{ph}}} \sim 
\frac{k_B T E_F}{(\hbar \omega_D)^2} \sim \frac{\hbar}{m_\star v_s^2} \frac{k_B T}{\hbar} \,.
\ee
Here $E_F \sim \hbar v_F k_F$. For the final step we estimated $\omega_D \sim v_s/a$ and $E_F \sim \hbar^2/(m_\star a^2)$. That is, we again assume that the Fermi surface occupies a significant fraction of the Brillouin zone. The relation (\ref{eq:planck}) follows from (\ref{eq:ratio}).

As in the textbook case for conventional electron-phonon materials such as Copper, a Planckian decay rate $\tau^{-1}_{\text{el}\to\text{ph}} \approx k_B T/\hbar$ is at the margin of validity of the simple scattering computation that we have performed \cite{P2}. At high temperatures this computation can be justified because the scattering becomes elastic and the energy spreading of the Fermi-Dirac distribution is irrelevant \cite{peierls1996quantum}. For temperatures of order the Debye temperature it has been argued by Prange and Kadanoff that the simple computations above remain valid due to the simplifications inherent in the electron-phonon problem, such as the hierarchy of the electron and phonon velocities \cite{PhysRev.134.A566}.

\section{Lorenz ratio for high T$_c$ materials}

\begin{figure}[h!]
    \centering
    \includegraphics[width=0.6\textwidth]{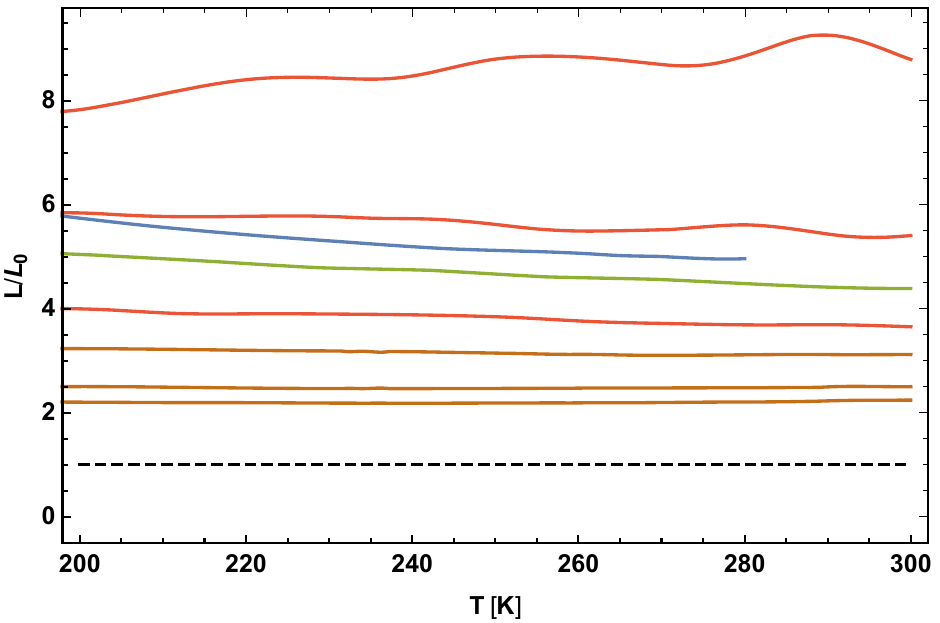}
    \caption{Lorenz ratio $L$ in the high T$_c$ materials shown in Fig.~\ref{fig:highTc} in the main text. Colors correspond to the colors in Fig.~\ref{fig:highTc}. The dashed line shows the Sommerfeld value $L = L_0$.}
    \label{fig:LorhighTc}
\end{figure}

Fig.~\ref{fig:LorhighTc} shows the Lorenz ratio $L =\kappa/(T \sigma)$ of thermal to electric conductivity for most of the high T$_c$ materials in Fig.~\ref{fig:highTc} in the main text. For the cases of YBCO, EBCO and Ba-122 the resistivity has been measured on the same samples as the thermal conductivity. For overdoped LSCO the resistivity is from similar samples, measured in \cite{PhysRevLett.93.267001}. Underdoped LSCO is not plotted here because we do not know the resistivity on sufficiently similar samples, but it is clear that $L/L_0$ will be much higher than for the overdoped LSCO shown.
The plot shows that $L > L_0$ over this temperature range, most dramatically so for the underdoped and optimally doped  compounds. Even the most overdoped LSCO material shown has $L \gtrsim 2.2 L_0$ so that more than half of the heat current is carried by phonons. We note in passing that phonons continue to dominate the thermal conductivity of these materials down to significantly lower temperatures than those shown.

The fact that $L/L_0$ is relatively constant over the temperature range shown is a consequence of the temperature dependence $\sigma \sim 1/T$ (this behavior is not precise away from optimal doping) while $\kappa$ is roughly independent of temperature. We emphasized in the main text, however, that $D_\text{th} = \kappa/c$ does depend on temperature --- in a way that directly reflects the underlying phonon lifetime. The approximate constancy of $\kappa$ appears as a coincidence from this point of view, due to a cancellation between the temperature dependence of $c$ and $\tau_\text{ph}$. In contrast, if the charge compressibility $\chi$ is constant (as would be expected for conventional electrons well below the Fermi energy), then the charge diffusivity $D_\text{ch} = \sigma/\chi \sim 1/T$, so that the temperature dependence of the electrical conductivity does directly reflect the underlying electronic quasiparticle lifetime.

\section{Phonons and electrons in doped semiconductors}
\label{sec:semiconductors}

In the main text we stated that heavily doped semiconductors can have phonon-dominated thermal transport while hosting degenerate charge carriers. This allows the `action-reaction' correspondence (\ref{eq:planck}) to be explored in a well-controlled setting. As noted in the main text, first principles transport studies in these materials have elucidated in detail the relative roles of disorder and electron-phonon scattering. Here we review transport data on heavily doped semiconductors, highlighting phenomenological similarities with the high T$_c$ materials.

We will illustrate the role of electron-phonon interactions on heat and charge transport with data from heavily doped Ge-Si alloys \cite{dis, ger, PhysRev.136.A1149} as well as heavily doped single silicon crystals \cite{doi:10.1002/aelm.201600171}. Mass disorder is well-known to be an important source of phonon scattering in the alloys \cite{PhysRev.125.44,PhysRev.131.1906,PhysRevLett.106.045901}. However, as explained in the main text, mass disorder does not efficiently scatter long wavelength phonons. Furthermore, one expects that the disorder introduced by doping is small compared to the disorder in the alloy itself while doping does control the number of charge carriers \cite{PhysRev.136.A1149}.
This helps to separate the effects of phonon scattering by disorder and by charge carriers.

\begin{figure}[ht!]
    \centering
    \includegraphics[width=\textwidth]{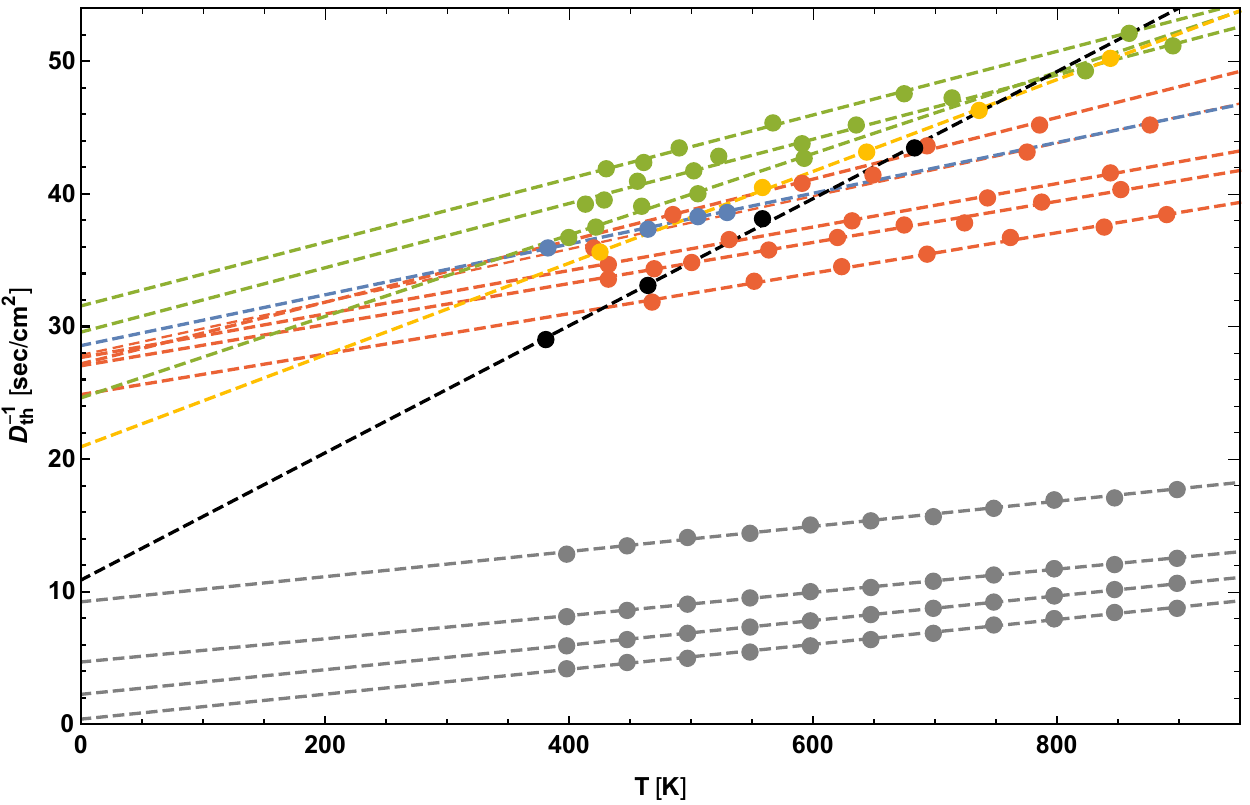}
    \caption{Inverse thermal diffusivity of heavily doped semiconductors. Dots are data points and dashed lines are fits to (\ref{eq:Dth}).
    Gray are P doped single crystal silicon samples with, from bottom to top, carrier density $n=2.3, 3.2, 3.55, 3.65 \times 10^{20}$ cm$^{-3}$ \cite{doi:10.1002/aelm.201600171}. The remainder are doped Ge-Si alloys \cite{dis, PhysRev.136.A1149}.
    Red have degenerate p-type carrier densities, from top to bottom, $n=1.8, 2.4, 3.5, 3.0, 2.1 \times 10^{20}$ cm$^{-3}$. The former three are 30\% Ge while the latter two are 15\% Ge. Yellow has mildly non-degenerate p-type carriers with $n=0.89\times 10^{20}$ cm$^{-3}$ and is 30\% Ge. Blue has degenerate n-type carriers with $n=2.7 \times 10^{20}$ cm$^{-3}$ and 15\% Ge. Green have mildly non-degenerate n-type carriers with, from top to bottom at lower temperatures, $n=1.5, 1.4, 0.67 \times 10^{20}$ cm$^{-3}$. These have 30\%, 20\% and 30\% Ge, respectively. Black is a far less doped sample (for comparison) with $n = 2.2 \times 10^{18}$ cm$^{-3}$ and 30\% Ge.}
    \label{fig:semi}
\end{figure}

Fig.~\ref{fig:semi} shows the inverse thermal diffusivity of various heavily doped samples over the temperature range of 400 to 900 K. The inverse thermal diffusivity is seen to have the additive form (\ref{eq:Dth}).

Fig.~\ref{fig:semi} is qualitatively similar to Fig.~\ref{fig:highTc} in the main text for high T$_c$ materials. Allowing for the lower effective mass of the semiconductors (while there is some mild doping and temperature dependence of $m_\star$, the discussion in \cite{doi:10.1063/1.349385} suggests that $m_\star \sim m_e$ in all cases), the more highly doped silicon crystals in particular have a comparable value of the dimensionless intercept $D_0^{-1} \times \hbar/m_\star$ as the high T$_c$ materials. In the alloys the intercept is larger due to mass disorder. In particular the intercept of the `undoped' alloy (black) is entirely due to disorder scattering. The increase in the intercept for the remaining doped alloys can then largely be associated to additional carrier scattering. It is seen that carrier and disorder scattering are thus roughly comparable, consistent with the estimate in the main text and with a first principles study \cite{doi:10.1063/1.5108836}.

\begin{figure}[h!]
    \centering
    \includegraphics[width=\textwidth]{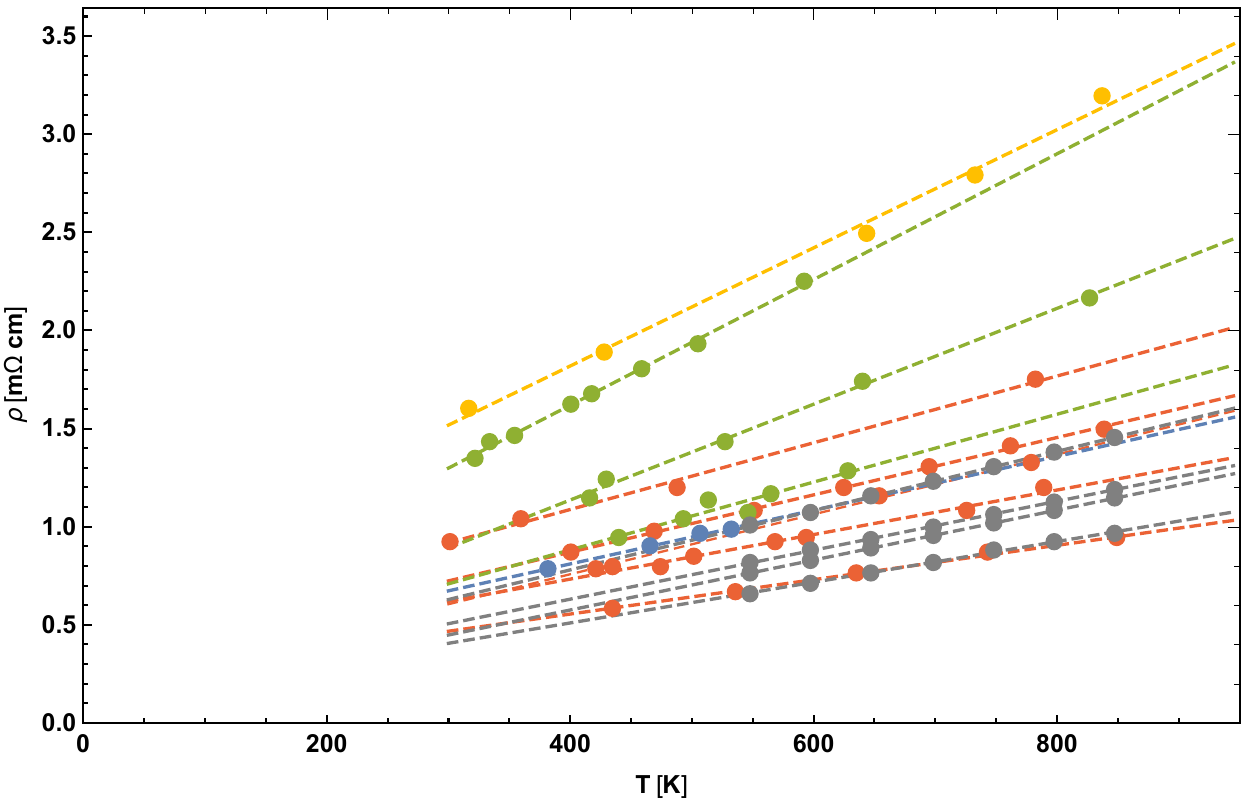}
    \caption{The resistivity $\rho$ in the heavily doped semiconductors shown in Fig.~\ref{fig:semi}. Colors correspond to the colors in Fig.~\ref{fig:semi}. Dots are data points and dashed lines show a linear fit.}
    \label{fig:rhosemi}
\end{figure}

All the samples shown in Fig.~\ref{fig:semi} --- except the `undoped' one, which was presented for comparison --- are either degenerate or only mildy non-degenerate over this temperature range, so that the resistivity is $T$-linear to a good approximation, with $\rho = \rho_0 + \rho_1 T$.
The resistivity of the same samples is shown in Fig.~\ref{fig:rhosemi}. Only the mildly non-degenerate p-type case (yellow) shows some curvature. The resistivities of the silicon crystal and the degenerate alloys are seen to be comparable. The less heavily doped alloys have a higher resistivity.

\begin{figure}[h!]
    \centering
    \includegraphics[width=0.48\textwidth]{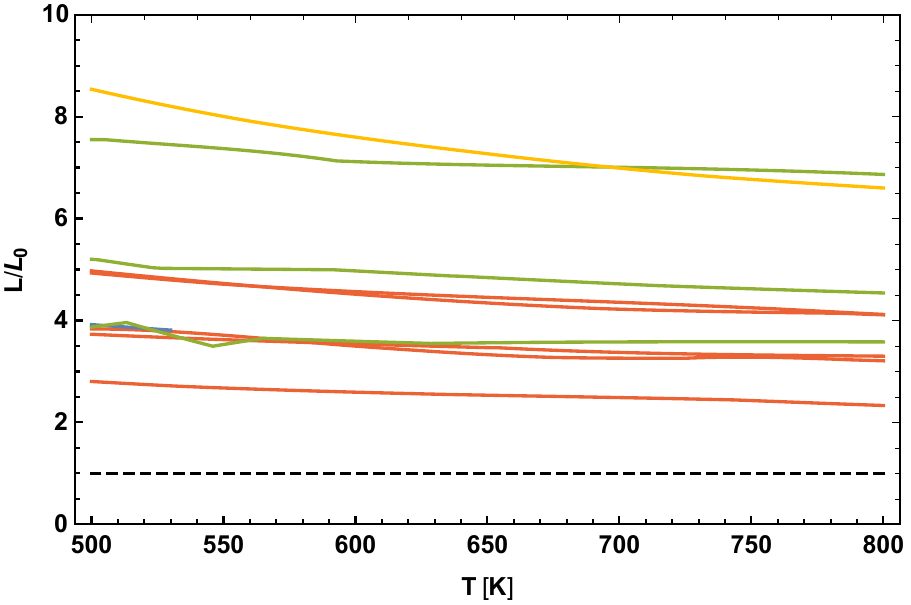}\includegraphics[width=0.48\textwidth]{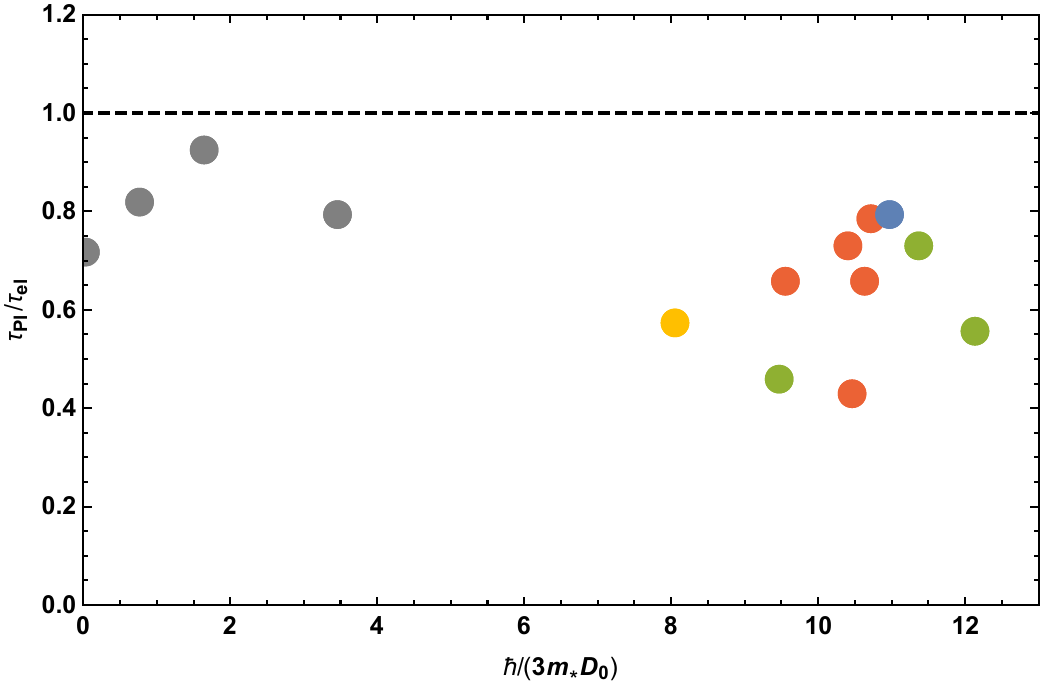}
    \caption{{\bf Left:} Lorenz ratio $L$ in the heavily doped semiconductors shown in Fig.~\ref{fig:semi} in the main text. Colors correspond to the colors in Fig.~\ref{fig:semi}. The doped single crystal silicon cases are not shown -- these have a much higher value of $L$. The dashed line shows the Sommerfeld value $L = L_0$. {\bf Right:} The inverse electronic lifetime in units of the Planckian time $\tau_\text{Pl} = \hbar/(k_B T)$ and the intercept $D_0^{-1}$ in units of $3 m_\star/\hbar$, for the heavily doped semiconductors in Fig.~\ref{fig:semi}. Colors correspond to the colors in Fig.~\ref{fig:semi}. The electronic lifetime is obtained from the resistivity as described in the text. }
    \label{fig:Lorsemi}
\end{figure}

The mean free path $\ell$ of the electrons in all cases is seen to obey $k_F \ell \sim 2 \pi$ over this temperature range, as determined from the resistivity by $k_F \ell = h/e^2 \times 3 \pi/(2 k_F) \times 1/\rho$ with the Fermi wavevector obtained from the carrier density as $n = k_F^3/(3 \pi^2)$. Quasiparticles are therefore only marginally well-defined, rather similarly to in the high T$_c$ materials considered in the main text. From the resistivity in Fig.~\ref{fig:rhosemi} we can also extract an inelastic electronic lifetime $\tau_{\text{el} \to \text{ph}}$ by fitting to $\rho = \rho_0 + \rho_1 T$ and then taking $\tau^{-1}_{\text{el} \to \text{ph}} = n e^2 \rho_1 T/m_\star$. As per the previous comments, i.e.~following \cite{doi:10.1063/1.349385}, we have taken $m_\star = m_e$ in all cases. In Fig.~\ref{fig:Lorsemi} we see that this leads to an electronic lifetime that is close to Planckian in all cases. This Planckian scattering of electrons by phonons is similar to conventional metals \cite{P2} and has been previously noted for degenerate semiconductors in \cite{PhysRevX.10.031025}.\footnote{Non-degenerate three dimensional charge carriers have resistivity $\rho \sim T^{3/2}$ due to a factor of the thermal velocity $v_\text{th} \sim \sqrt{T}$ appearing instead of $v_F$. We see here that mildly non-degenerate charge carriers in semiconcuctors have numerically Planckian decay rates due to phonon scattering.} In Fig.~\ref{fig:Lorsemi} we furthermore verify that $L > 2 L_0$ over this temperature range, so that phonons dominate heat transport.

In Fig.~\ref{fig:Lorsemi} we also show the dimensionless diffusivity intercept in units $D_0^{-1} \times \hbar/(3 m_\star)$ for these samples. The factor of $d=3$ here is to account for the factor in the diffusivity (\ref{eq:diff}).
For the alloys we have noted that about half of the value of $D_0^{-1}$ is due to disorder scattering. Disorder scattering is less important for the single-crystal silicon samples. From the silicon data alone it is not clear to what extent the increase of $D_0^{-1}$ away from zero with doping is due to the disorder inherent in the doping or due to the increasing size of the Fermi surface with doping. Regarding the size of the Fermi surface, what matters is whether a given phonon can excite a particle-hole pair close to the Fermi surface. This requires the phonon wavevector to be of order or smaller than $k_F$. A detailed discussion of how many of the heat-carrying phonons obey this criterion at a given temperature and carrier density depends on band structure details, see e.g. \cite{PhysRevLett.114.115901} for doped silicon (showing that intervalley scattering processes can also be important).  The first principles study \cite{PhysRevLett.114.115901} as well as fits to the Callaway model \cite{doi:10.1002/aelm.201600171} suggest that scattering of phonons by carriers is significant for the samples considered here.

Finally, we note that significant scattering of heat-carrying phonons by charge carriers has also been studied in other doped semiconductors, e.g. \cite{Fu2015, manip}. We have focused on the materials above because they show a $T$-linear resistivity, corresponding to degenerate or only mildly non-degenerate electrons scattering from phonons, and furthermore have large Lorenz ratio. An offset in the high temperature inverse thermal diffusivity is also visible in doped strontium titanate \cite{PhysRevLett.120.125901}.

\providecommand{\href}[2]{#2}\begingroup\raggedright\endgroup

\end{document}